\documentclass[a4paper,fleqn,usenatbib,useAMS]{mnras}

\usepackage[T1]{fontenc}

\DeclareRobustCommand{\VAN}[3]{#2}
\let\VANthebibliography\thebibliography
\def\thebibliography{\DeclareRobustCommand{\VAN}[3]{##3}\VANthebibliography}

\usepackage{color}
\usepackage{graphicx}
\usepackage{enumitem}
\usepackage{amsmath}	
\usepackage{amssymb}	
\usepackage{multicol}        
\usepackage{bm}		
\usepackage{pdflscape}	
\usepackage{subcaption}
\usepackage{comment}

\definecolor{darkgreen}{rgb}{0.0,0.5,0.0}

\definecolor{grey}{rgb}{0.4,0.5,0.6}
\definecolor{trolleygrey}{rgb}{0.5, 0.5, 0.5}

\newcommand*\diff{\mathop{}\!\mathrm{d}}

\usepackage[T1]{fontenc}
\usepackage{ae,aecompl}
\usepackage{enumitem}
\usepackage{colortbl}

\usepackage{newtxtext,newtxmath}

\usepackage{etoolbox}

\title{Absence of obvious tidal tails around the globular cluster NGC~6397} 

\author[P. Boldrini and E. Vitral ]{Pierre Boldrini$^{1}$\thanks{Contact e-mail: \href{mailto:boldrini@iap.fr}{boldrini@iap.fr}} and {Eduardo Vitral$^{1}$\thanks{Contact e-mail: \href{mailto:vitral@iap.fr}{vitral@iap.fr}}}
\\
$^{1}$Sorbonne Universit\'e, CNRS, UMR 7095, Institut d'Astrophysique de Paris, 98 bis bd Arago, 75014 Paris, France}
\date{In original form 2020 September 1.}

\pubyear{2020}

\begin{document}
\label{firstpage}
\pagerange{\pageref{firstpage}--\pageref{lastpage}}
\maketitle

\begin{abstract}

In this work, we use $N-$body simulations performed on GPU to trace the past 10 Gyr dynamical history of a globular cluster (GC) similar to NGC~6397 in the tidal field of a Milky Way-like galaxy and we compare our simulated GCs with data from the third Gaia early data release. Our simulations predict, in contrast to what is deduced from the data, that such a cluster should present strong and extended tidal tails by more than 6 Gyr ago (right after the first third of its life), exceeding 1 kpc of length, and should be roughly disrupted by current time. We analyzed each of our initial conditions, such as initial mass and density parameters, as well as the  dark matter shape, and we argue that the most likely reason for such discrepancy between the data and our simulations is related to the fact that we consider a purely baryonic cluster in the beginning of each model we test. We discuss that if our globular cluster was initially embedded in a dark matter minihalo, the latter could act as a protecting envelope, which prevents the tidal stripping of the luminous matter, while being itself gradually disrupted and removed in the course of the cluster evolution. This could explain why an insignificant amount of dark matter is required to describe the velocity dispersion in NGC~6397, up to at least a few half-mass radii.

\end{abstract}

\begin{keywords}
galaxy dynamics -- (Galaxy:) globular clusters: individual: NGC 6397 -- Milky Way --  methods: orbital integrations -- $N-$body simulations
\end{keywords}




\section{Introduction}

Globular clusters (GCs) are among the oldest known stellar systems in astrophysics. These sources are spherically shaped collections of stars, often associated with very poor metallicities and usually occupy the outer regions of their host galaxy, within its dark matter (DM) halo (see \citealt{Gratton+19} for a complete review). The Milky Way (MW) itself hosts more then 150 GCs (\citealt{Harris10}), some of which occupy relatively close positions with respect to the Galaxy centre (\citealt{GaiaCollaboration+18H}).

Although GCs might seem as simple sources at first, they are subject to various enigmas that are yet not completely understood. For example, there is no clear consensus pertaining to their formation mechanisms (e.g. \citealt{Forbes&Bridges&Bridges10}): they could have been formed in smaller galaxies that later merged to become their present host galaxy (e.g. \citealt{Searle&Zinn78}, \citealt{Bullock&Johnston05}, \citealt*{Abadi+06} and \citealt*{Penarrubia+09}), as well as they could have been formed in-situ. There is also the possibility that they were formed in DM minihalos in the early Universe \citep{Peebles&Dicke68,Peebles84}, but we do not consider this dark component in our simulations.

Moreover, GCs serve as laboratories to many astrophysical studies such as the motion and internal kinematics of spherical systems (e.g. \citealt{Baumgardt2019}, \citealt{Vasiliev2019}),  intermediate-mass black holes (\citealt*{Greene+19}) and multiple stellar populations (\citealt{Carretta+09}). Therefore, it is important to understand the impact of dynamics on the long-term evolution of GCs so that one can trace back their history and better model the underlying physics of these complex sources.

In this work, we examine whether stellar dynamics is sufficient to explain the survival of the GCs under the effect of strong tidal fields of the MW over the course of their evolution via $N-$body simulations. To perform such an analysis, we model the nearby GC NGC~6397, the second closest GC to our Sun at only 2.39 kpc away (\citealt{Brown+18}), with a pericentre of some few kpc (1.7 -- 2.9 kpc, \citealt{GaiaCollaboration+18H}), thus undergoing considerable tidal interactions with the MW. NGC~6397 had its dynamics broadly analyzed (e.g. \citealt{Heyl+12}, \citealt{Watkins+15a}, \citealt{Watkins+15b}, \citealt{Husser+16}, \citealt{Baumgardt2019} and \citealt{Vitral&Mamon21}) and was observed many times, by different space missions such as the Gaia Space Telescope \citep{GaiaCollaboration+18H} and the Hubble Space Telescope \citep{Bellini+14}.

Our $N-$body simulations performed on GPU indicate that, when assuming a purely baryonic source and a cuspy DM density for the MW, strong tidal effects should have started disrupting this cluster after the first third of its life due to the dense background of DM and stars in our Galaxy. We tested different scenarios such as the presence of a central black hole, as well as different density slopes and characteristic radii. Then, we compare our simulation results to observations from \textsc{Gaia EDR3}. Our paper is organized as follows: In Section 2, we describe the most recent data concerning NGC 6397 derived from observations. Section 3 provides a description of the $N-$body modelling. In Section 4, we outline details of our numerical simulations and method. In Section 5, we discuss our results and their interpretation. In Sections 6 and 7, we outline and draw our conclusions.


\section{Observations of NGC 6397}

Given its proximity, the dynamics of NGC~6397 was broadly observed during the course of recent years: \cite{Bellini+14} performed deep observations of the cluster's core using the Hubble Space Telescope (HST) and provided thousands of proper motions, which were later complemented by the proper motions from the data releases of the \text{Gaia astrometric mission} (hereafter, \textsc{Gaia}), better probing its outskirts (\citealt{GaiaCollaboration+18H}, \citealt{GaiaCollaboration+20}). In addition, line-of-sight velocities were also measured in various studies such as \cite{Husser+16} and \cite{Lovisi+12}.

\subsection{Dynamics}

With this vast amount of velocity data, the dynamical analysis of this GC is also the subject of many studies: \cite{Heyl+12} analyzed main sequence stars along with white dwarfs from HST observations and measured a mass of $1.1 \times 10^5$ M$_{\odot}$ for a photometric distance of 2.53 pc (even though they estimate a kinematic distance of 2.2 kpc). They also compared the velocity dispersion and effective radii in different magnitude bins and assigned a considerable mass segregation to this GC. 
In addition, \cite{Husser+16} used MUSE spectrograph data to measure a mean line-of-sight (LOS) velocity of thousands of stars from this GC, to which \cite{Kamann+16} later fitted a 600 M$_{\odot}$ central black hole.

More recently, \cite{Vitral&Mamon21} combined data from HST, \textsc{Gaia DR2} and the Multi Unit Spectroscopic Explorer (MUSE) to perform a Bayesian mass-modeling analysis of this cluster through the handling of the Jeans equation.
They found strong evidence for isotropy up to 8 arcmin (i.e. $\sim5.6$ pc) from the cluster's centre and used a distance of 2.39 kpc (\citealt{Brown+18}) to estimate a total luminous mass of $1.2 \times 10^5$ M$_{\odot}$. They provided important constraints on the surface density profile by modelling the inner cusp of this post core-collapse cluster with a S\'ersic model (\citealt{Sersic63}) and also found strong mass segregation by jointly modeling stellar populations with different mean mass. Finally, although this work managed to fit an intermediate-mass black hole of mass $511$ M$_{\odot}$ located in this cluster's centre, the authors ruled it out in favour of an unresolved inner sub-cluster of stellar remnants, after analyzing both the statistics and the physics involved in both scenarios. 

Using the orbital radius from \textsc{Gaia DR2} data \citep{Vasiliev2019} and the mass from \cite{Vitral&Mamon21}, we calculate the theoretical tidal radius of NGC~6397 using the formula of \cite{Bertin2008}. 
All these analyses provide us with strong constraints regarding the choice of parameters in our simulations (see Table~\ref{tab: observation-ngc}), and thus allow us to interpret our conclusions with better confidence.

\subsection{Tails}

The strong tidal field of the MW could, in principle, form extended tidal tails in NGC~6397. \cite*{Leon+00} found, while looking for shock imprints, that this GC presented few over-densities (tidal tails) at larger radii, which they were not able to robustly distinguish from dust extinction and therefore classified them as unreliable measurements. Recently, \cite{Kundu+20} claimed to have identified tidal tails beyond the cluster tidal radius, which could be related to shocks and tidal disruption.
\citeauthor{Kundu+20} argue that it was beyond the scope of their work to derive a precise fit of the proper motion distribution, and that an idea of the cluster intrinsic dispersion was sufficient to select tidal stars. However, at regions distant from the cluster centre, such as the ones probed in their work, the contamination of MW field stars can be increasingly problematic, especially when targeting a small number of stars (120 tidal stars).

In a recent work, \cite{Ibata+20} performed a thorough search for stellar streams in many MW GCs and successfully identified an irregular stellar stream around NGC 6397, extending 18$^{\rm o}$ on the sky, which would mean, disregarding projection effects, a tail of roughly 750 pc.
In the following sections, we try to detect even more extended tidal tails in NGC~6397.

\section{Numerical modelling}

In this section, we comment some of the previous attempts to numerically model NGC~6397 and then present the models for the MW and NGC 6397, that provide the initial conditions for our simulations.

\begin{table}
\begin{center}
\begin{tabular}{ccccccccccc}
 \hline
   NGC 6397 & Stellar mass & $1.17^{+0.12}_{-0.16} \times 10^{5}M_{\sun}$ \\
   & Sérsic radius & $3.14 \pm 0.25$ pc \\
   & Sérsic index & $3.26 \pm 0.23$ \\
   & Orbital radius & $5.91 \pm 0.06$ kpc \\
   & Tidal radius & $59.9 \pm 3.2$ pc \\
    \hline
\end{tabular}
\caption{{\it Observations of NGC 6397:} We list the main information used concerning the mass, density profile, and orbital parameters of NGC 6397. The mass, S\'ersic radius and index were obtained from \protect\cite{Vitral&Mamon21}, the orbital radius was taken from \textsc{Gaia DR2} data \protect\citep{Vasiliev2019} and the tidal radius was derived from this set of values using the relation from \protect\cite{Bertin2008}.}
\label{tab: observation-ngc}
\end{center}
\end{table}

\subsection{Previous modeling}

NGC~6397 has been modelled before by means of both $N-$body and Monte Carlo simulations. \cite{Heggie&Giersz&Giersz09} placed this cluster in a tidal field and performed Monte Carlo simulations in order to study the evolution of the central regions of the cluster, through the course of 1 Gyr, and provide a mass loss rate of $\diff \ln{M} / \diff t = -1.35 \times 10^{-4}$, with $t$ in Myr. They do not mention, however, the characteristics of the tidal field used. \cite{Baumgardt2003} performed robust simulations of GCs in tidal fields, and assigned a high dissolution time for NGC~6397 (24.3 Gyr), but this was done considering a constant Milky Way potential, which could affect the results of GCs placed in inner orbits.

Recently, other GCs have been modelled, and their tidal tail properties could be better accessed. For instance, \cite{Wan+21} analyzed the dynamics of NGC 3201 and matched simulations to observations to better understand the properties of its tidal tails, and \cite{Gieles+21} simulated Palomar 5 in a realistic Milky Way potential and showed that the presence of visible tidal tails is directly related to an important population of stellar-mass black holes (see their Figure 4). In fact, since NGC~6397 is a core-collapse cluster, believed to be almost depleted of stellar mass black holes (e.g. \citealt{Kremer+20}), one could, in principle, expect that it would not produce prominent stellar streams.

\subsection{NGC 6397}

GCs are often modelled by a King profile (\citealt{King66}), which displays a flat core. Nevertheless, we have chosen to adopt a S\'ersic profile because it should be more resilient to the tidal stripping by the MW as the GC potential is deeper relative to that from a King profile. Besides, it allows us to test how much the inner slope of the cluster could impact its survival in such strong tidal field. Our GC assumes then the analytical approximation of \cite{Prugniel&Simien97}:

\begin{equation}
    \rho(r)=\rho_{0}\left(\frac{r}{R_{\mathrm{e}}}\right)^{-p_{n}} \exp \left[-b_{n}\left(\frac{r}{R_{\mathrm{e}}}\right)^{1/n}\right] \ ,
    \label{eq: prugniel}
\end{equation}
where $R_{\mathrm{e}}$ is the effective radius containing half of the projected luminosity, $n$ is the S\'ersic index, $b_n$ is a function of $n$, which can be approximated with the precise relation from \cite{Ciotti&Bertin99} and $p_{n}$ is also a function depending only on $n$, well constrained by \cite{LimaNeto+99}. Since the GC is thought to have suffered from tidal interactions with the MW, we suppose that this GC was initially much more massive in the past. We assigned initial masses up to five and ten times the current cluster mass \citep{Baumgardt2019}, as can be seen in Table~\ref{tab: observation-ngc}. For all the simulations, we chose an initial S\'ersic radius $R_{\mathrm{e}}\leq1$ pc, lower than the observed radius (see Table~\ref{tab: observation-ngc}), because it is susceptible to increase through dynamical processes such as mass loss. Moreover, \cite*{Lamers+10} suggested that most of GCs probably formed with half-mass radii around 1 pc. We also tested a GC density profile with a steeper inner slope, by increasing the initial S\'ersic index, up to a value of 6. Steeper density profiles should help the cluster to keep its form for longer timescales than cored profiles following a King model. 

\subsection{Milky Way}

\begin{table}
\begin{center}
\begin{tabular}{ccccccccccc}
 \hline
   Milky Way & Profile & a & $r_{200}$ &  Mass \\
     & & [kpc] & [kpc] & [$10^{10}M_{\sun}$] \\
    \hline
   Halo & NFW & 13.6 & 300 & 60 \\
   Bulge & Hernquist & 0.64 & - & 0.46 \\
   Disk & Exponential disk & $R_{\mathrm{d}}=2.8$ & - & 5 \\
   & & $z_{\mathrm{0}}=0.39$& \\
    \hline
\end{tabular}
\caption{{\it Milky Way model:} From left to right, the columns give for each MW component: the density profile; the scale length; the virial radius, and the mass. The mass model of MW that we use is based on \protect\cite{Widrow+08}. It consists of a Hernquist bulge \citep{Hernquist90} and stellar exponential disk which are embedded inside an NFW DM halo \citep{Navarro+96}}
\label{tab: galaxy-params}
\end{center}
\end{table}

\begin{figure}
\centering
\includegraphics[width=0.47\textwidth]{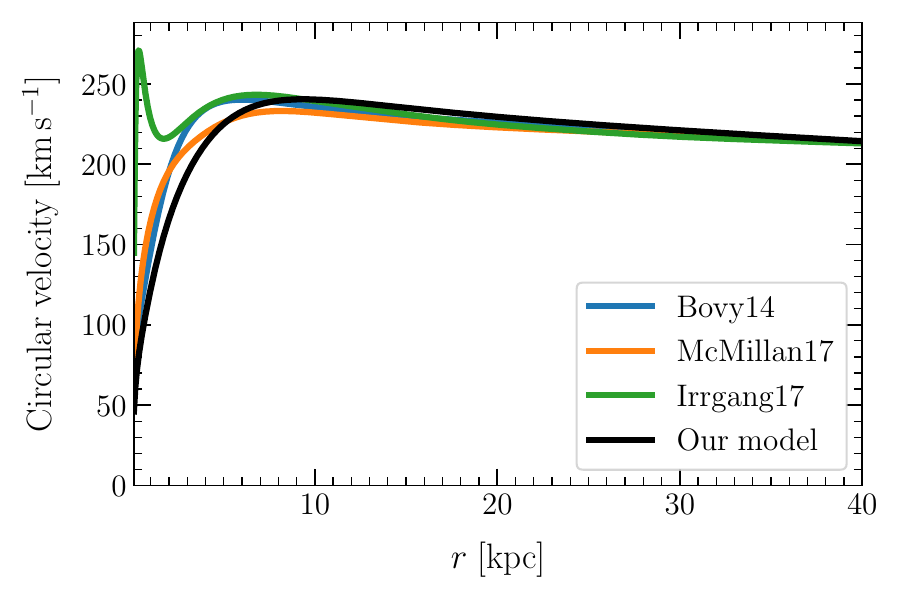}
\caption{{\it Comparison between Milky Way mass models:} Total circular velocity as a function of the radius for different MW mass models \citep{Bovy15,McMillan17,Irrgang+13}. Our model has the lowest circular velocity in the MW inner region (below 10 kpc). We have therefore chosen this mass model in order to maximize the lifetime of the cluster.}
\label{fig: circ-velocity}
\end{figure}

Our Galaxy is modelled as a static potential. The mass model of the MW that we use is based on \cite{Widrow+08}. It consists of a Hernquist bulge \citep{Hernquist90} and stellar exponential disk which are embedded inside an NFW DM halo \citep{Navarro+96}. The MW parameters are summarized in Table~\ref{tab: galaxy-params}. Figure~\ref{fig: circ-velocity} compares our MW model with other MW potentials from the literature \citep{Bovy15,McMillan17,Irrgang+13}. 

The strength of the tidal stripping depends directly on the MW mass enclosed within the orbit. It can be seen that our model has the lowest circular velocity in the MW inner region (below 10 kpc) (see Figure~\ref{fig: circ-velocity}). We have therefore chosen this mass model in order to maximize the lifetime of the cluster. In fact, if we choose the potential of \cite{Irrgang+13}, as in \cite{GaiaCollaboration+18H}, we expect to have a stronger tidal field in the central region of the Galaxy and thus a shorter lifetime for the cluster. 

The major accretion event of the MW has occurred between 9 and 11 Gyr ago \citep{DiMatteo+19}. Moreover, the mass of the simulated MW-like halo from Via Lactea-1 reaches its asymptotic value 9-10 Gyr ago \citep{Diemand+07,Lux+10}. As a consequence, we establish that our model describes well the Galaxy until 9-10 Gyr backward in time. 

Furthermore, our pure $N$-body simulations do not include mass loss from stellar evolution, which acts within the first Gyr of GC evolution \citep{Baumgardt2003}. As NGC 6397 is very old (12.87 Gyr,  \citealt{MarinFranch+09}), we expect that the absence of this physical process will not strongly affect the GC dynamics in our simulations, which start 10 Gyr ago.

\subsection{Initial conditions}

We take the present-day phase-space position of NGC~6397 derived from \textsc{Gaia DR2} data \citep{Vasiliev2019} and integrate them backward in time for 10 Gyr in our MW potential. Orbit integration is performed with a time step of 10 Myr using the publicly available code \textsc{Galpy} by applying dynamical friction to this cluster \citep{Bovy15}. The time to change the apocentre substantially due to dynamical friction is proportional to M$_{\mathrm{enclosed}}$/M$_{\mathrm{GC}}$ $\times$ t$_{\mathrm{dyn}}$ where M$_{\mathrm{enclosed}}$ is the host galaxy mass enclosed within the orbit and t$_{\mathrm{dyn}}$ is the orbital time \citep{Binney2008}. As the mass ratio between the MW enclosed mass and the GC mass is large, dynamical friction is inefficient over our timescale. This is the rationale underlying our assumption that the GC mass remains constant during the orbital integrations. The orbit integrations were done for NGC~6397 with its current mass undergoing dynamical friction (see details in Table~\ref{tab: observation-ngc}). Besides, assuming a static potential for the MW galaxy in our simulations is still valid as dynamical friction is then negligible compared to tidal effects caused by the Galaxy.

\begin{figure}
\centering
\includegraphics[width=0.47\textwidth]{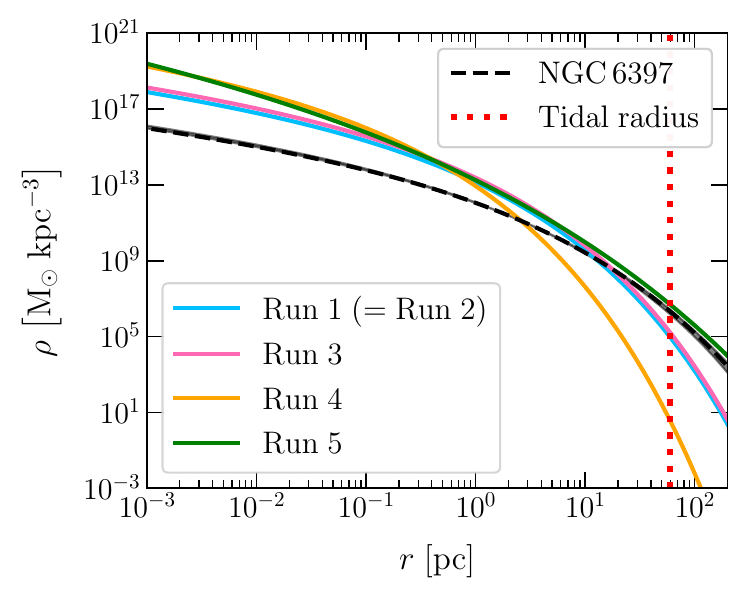}
\caption{{\it Initial stellar distribution:} Initial S\'ersic profile for our simulated GC for the different runs (see Table~\ref{tab: n-body-models}). \protect\cite{Vitral&Mamon21} determined the fit (dashed black line) to the observed stellar distribution of NGC~6397 that we use in this work (see Table~\ref{tab: observation-ngc}). The observed tidal radius is marked by the vertical dashed red line.}
\label{fig: dens-0}
\end{figure}

\section{Pure N-body simulations on GPU}
\label{sec: Nbody}

\begin{figure}
\centering
\includegraphics[width=\hsize]{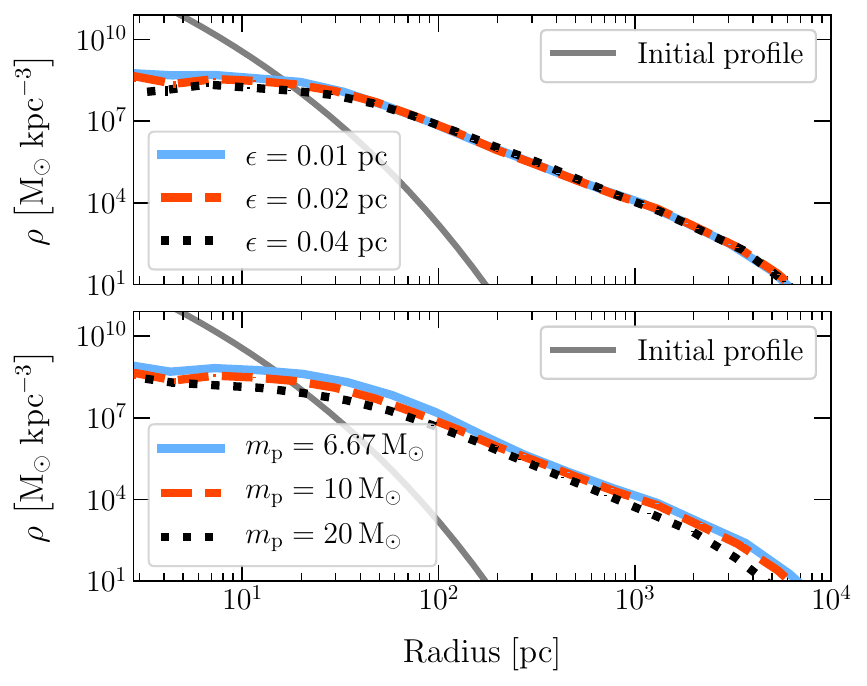}
\caption{\textit{Robustness of the simulations.} \textit{Top:} Density profile of our simulated GC at the beginning of the simulation (in grey) and at their disruption time (other colors) for different softening lengths (see Table~\ref{tab: n-body-models}). Our simulations are well converged for softening lengths ranging from $\epsilon=$ 0.01 and 0.04 pc. \textit{Bottom:} Density profile of our simulated GC at the beginning of the simulation (in grey) and at their disruption time (other colors) for different stellar mass resolution. We see that the evolution of the spatial distribution of stars for GCs with different $m_{\mathrm{p}}$ is very similar but on different time-scales in our pure $N$-body simulations (see Table~\ref{tab: n-body-models}). The error bars display Poisson errors.}
\label{fig: robustness}
\end{figure}

\begin{table*}
\begin{center}
\begin{tabular}{cccccccccc}

Simulation & $M_{\mathrm{GC}}$ & $M_{\mathrm{BH}}$ & $R_{\mathrm{e}}$ & n & N & $\epsilon$ & $T_{\mathrm{d}}^{\mathrm{sim}}$ & $T_{\mathrm{d}}^{\mathrm{c}}$ & $T_{\mathrm{tails}}^{\mathrm{c}}$\\
 & [M$_{\sun}$] & [M$_{\sun}$] & [pc] & &   & [pc] & [Gyr] & [Gyr] &  [Gyr]\\
    \hline
    Run 1 & 5.75$\times10^{5}$ & 0 & 1.0 & 3.3 & 57500 & 0.02 & 1.18& 8.48 & 3.41\\
    \hline
    Run 2 & 5.75$\times10^{5}$ & $10^{3}$ & 1.0 & 3.3 & 57500 & 0.02 & 1.11 & 7.98 & 3.14\\
    Run 3 & 1$\times10^{6}$ & 0 & 1.0 & 3.3 & 1$\times10^{5}$ & 0.02 & 1.38 &  10.1 & 3.43\\
    Run 4 & 5.75$\times10^{5}$ & 0 & 0.2 & 3.3 & 57500 & 0.005 & 1.49 & 10.71 & 3.59\\ 
    Run 5 & 1$\times10^{6}$ & 0 & 1.0 & 6 & 1$\times10^{5}$ & 0.02 & 1.68 & 12.28 & 3.56\\
    \hline
    Run A1 & 5.75$\times10^{5}$ & 0 & 1.0 & 3.3 & 57500 & 0.01 & 1.18 & 8.48  &  3.43\\
    Run A2 & 5.75$\times10^{5}$ & 0 & 1.0 & 3.3 & 57500 & 0.04 & 1.18 & 8.48  &  3.41\\
    Run B1 & 5.75$\times10^{5}$ & 0 & 1.0 & 3.3 & 86250 & 0.02 & 1.56 & 8.42  &  3.41\\
    Run B2 & 5.75$\times10^{5}$ & 0 & 1.0 & 3.3 & 28750 & 0.02 & 0.76 & 8.86  &  3.06\\
    Run B3 & 5.75$\times10^{5}$ & 0 & 1.0 & 3.3 & 5750 & 0.02  & 0.26 & 8.93  &  -\\
    \hline
\end{tabular}
\caption{Simulation parameters for all the scenarios. From left to right, the columns indicate: the GC mass; the central BH mass; the Sérsic effective radius; the Sérsic index; the particle number; the softening length; the disruption time in the simulation; the corrected disruption time using equation~\eqref{eq: correction}; the corrected tail formation time using equation~\eqref{eq: correction}. Because of an insufficient number of particles in Run B3, we do not observe any tidal arms.}
\label{tab: n-body-models}
\end{center}
\end{table*}

To generate the live GC, we use the initial condition code \textsc{magi} \citep{Miki&Umemura&Umemura18}. Adopting a distribution-function-based method, it ensures that the final realization of the cluster is in dynamical equilibrium \citep{Miki&Umemura&Umemura18}. We perform our simulations with the high performance collisionless $N$-body code \textsc{gothic} \citep{Miki&Umemura&Umemura17}. This gravitational octree code runs entirely on GPU and is accelerated by the use of hierarchical time steps in which a group of particles has the same time step \citep{Miki&Umemura&Umemura17}. We evolve the MW-GC system over 10 Gyr for all runs. The GC initial position and velocity are derived from the orbital integration of NGC~6397 in our MW potential backward in time over 10 Gyr (see Table~\ref{tab: galaxy-params}). 

The softening value was estimated using the following criterion: $\epsilon\sim R_{\mathrm{e}}/N^{1/3}$, where $N$ and $R_{\mathrm{e}}$ are the number of particles and the S\'ersic radius of the GC, respectively. As in \textsc{Gadget-2} \citep{Springel05}, our GPU $N$-body code is accelerated by the use of a block time step (see details in \citealt{Miki&Umemura&Umemura17}). The time step is proportional to the time-step parameter $\eta$ whose the default value is 0.5. Most of the runs reported here were set with a GC mass resolution of 10 M$_{\sun}$ and a softening length of 0.02 pc. We have repeated runs by varying the softening length in order to ensure that our simulations do not suffer from numerical noise. The impact of this parameter on the evolution of the stellar density profile is shown in Figure~\ref{fig: robustness} for a GC assuming an initial S\'ersic radius of 1 pc and index of 3.3. According to the top panel of Figure~\ref{fig: robustness}, our simulations are well converged for softening lengths ranging from $\epsilon=$ 0.01 and 0.04 pc. At the disruption time in the simulation $T_{\mathrm{d}}^{\mathrm{sim}}$ (see Table~\ref{tab: n-body-models}), the stellar density profiles confirm that it is sufficient to consider a softening length of 0.02 pc for our study (see Figure~\ref{fig: robustness}). 

In our pure $N$-body simulations, GCs dissolve as a result of two-body relaxation and external tidal shocks. However, the choice of mass resolution can significantly affect the relaxation time and therefore the lifetime of GCs in the simulation. This numerical effect is more pronounced in strong tidal fields such as in the MW inner region. We assess the impact of the mass resolution $m_{\mathrm{p}}$ on the density profile of GCs at their disruption times $T_{\mathrm{d}}^{\mathrm{sim}}$ on the bottom panel of Figure~\ref{fig: robustness}. The disruption times for all tests confirmed that the lifetime of GCs in the simulation depends significantly on the mass resolution, as they all ranged around 1 Gyr (see Table~\ref{tab: n-body-models}). Nevertheless, no strong variation was seen in the stellar density profile of GCs at their corresponding disruption time (see bottom panel of Figure~\ref{fig: robustness}). We observed that the evolution of the spatial distribution of stars for GCs with different $m_{\mathrm{p}}$ is very similar but on different time-scales in our pure $N$-body simulations. Indeed, increasing mass resolution accelerates the two-body relaxation and the MW tidal stripping. More precisely, it was shown that the disruption time of a GC behaves as $[N/\ln(0.02 \, N)]^{0.82}$ where $N$ is the number of particles \citep{Baumgardt2003}. Thus, it is possible to estimate from lower resolution simulations an approximation of the disruption time via the following correction factor: 

\begin{equation}
    C_{\mathrm{t}}=\left[\frac{m_{\mathrm{p}} \, \ln(0.02 \, M_{\mathrm{GC}}/m_{\mathrm{p}})}{\Bar{m} \, \ln(0.02 \, M_{\mathrm{GC}}/\Bar{m})}\right]^{0.82},
\label{eq: correction}
\end{equation}
where $\Bar{m}=$ 0.65 M$_{\sun}$ is the mean mass of a star in GCs \citep{Baumgardt2019}. As realistic simulations with $m_{\mathrm{p}}=$ 1 M$_{\sun}$ GC are extremely computationally costly, we investigate the dynamics of NGC 6397 by simulating GCs with $m_{\mathrm{p}}=$ 10 M$_{\sun}$ in different scenarios (see Table~\ref{tab: n-body-models}) and derived the disruption time $T_{\mathrm{d}}^{\mathrm{sim}}$. We assume that the formation of tails occurs when the stellar arms are extended over a distance of at least 1 kpc and this phenomenon is characterized by the tail formation time $T_{\mathrm{tails}}^{\mathrm{sim}}$. Then, we applied a correction to these characteristic times using equation~\eqref{eq: correction} and we obtained the corrected disruption time $T_{\mathrm{d}}^{\mathrm{c}}$ and the corrected tail formation time $T_{\mathrm{tails}}^{\mathrm{c}}$. 

According to Table~\ref{tab: n-body-models}, simulations with different $m_{\mathrm{p}}$ show that the corrected value $T_{\mathrm{d}}^{\mathrm{c}}$ seems to converge within 1 per cent for $m_{\mathrm{p}}\leq$ 10 M$_{\sun}$. Because of an insufficient number of particles in Run B3, we do not observe any tidal arms. We insist on the fact that it is necessary to have a mass resolution $\lesssim$ 10 M$_{\sun}$ to correctly resolve the dynamics in the outer regions of the GC such as in the tidal tails. Indeed, we observe some deviations around $T_{\mathrm{d}}^{\mathrm{c}}$ progressively as we depart from our mass resolution, towards higher values of $m_{\mathrm{p}}$ (see Run B2 and B3 in Table~\ref{tab: n-body-models}).

\begin{figure*}
\centering
\includegraphics[width=\hsize]{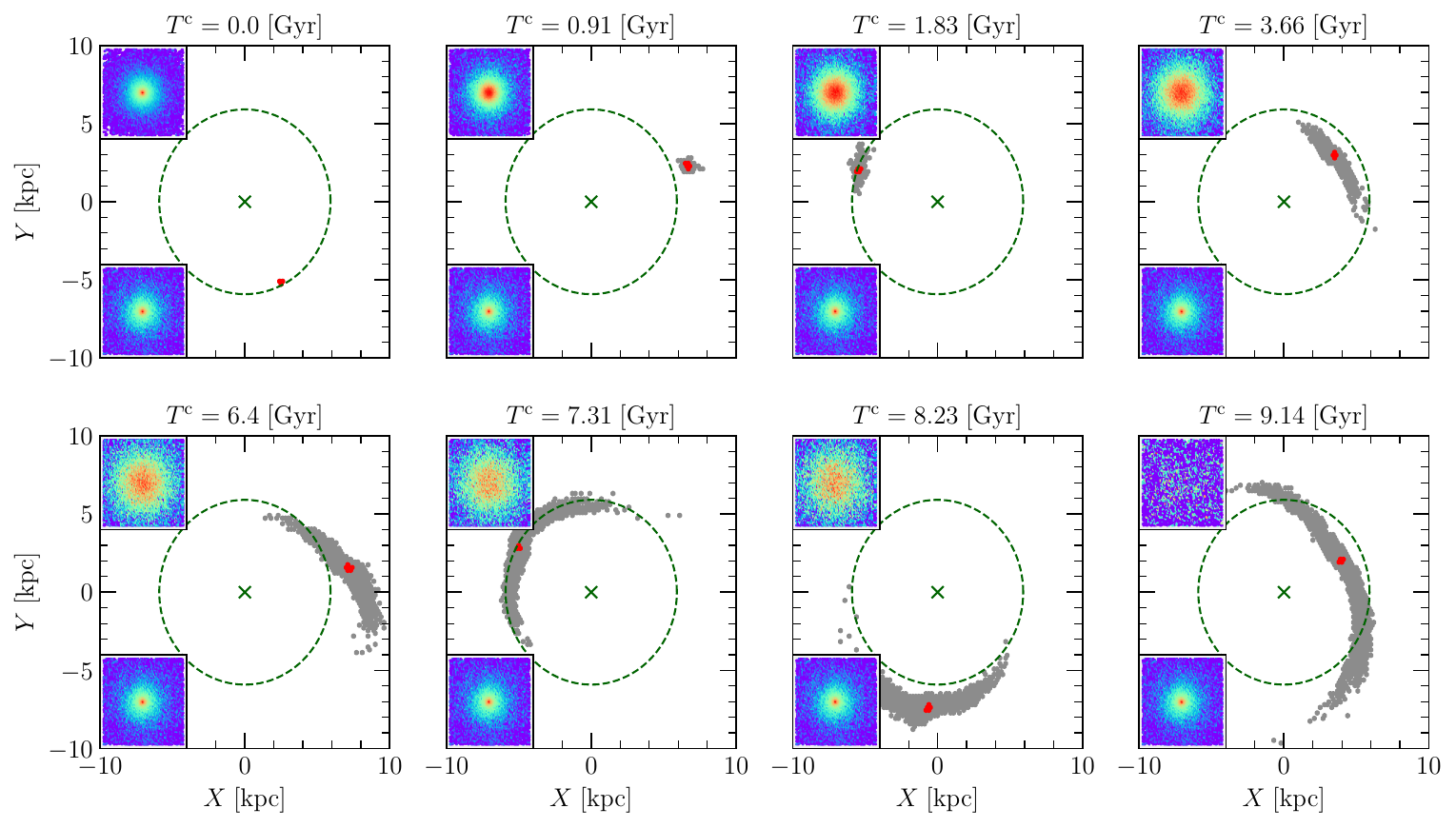}
\caption{{\it Formation of tidal tails:} Snapshots of the GC at different times for Run 3. Bounded and unbounded particles are represented in red and grey, respectively. The subplots on the left are a zoomed-in view of the bounded particles at each time, over a region up to $\sim$6 pc from its centre, with the top plots representing the evolving GC in our simulation and the bottom plots depicting a NGC~6397-like GC using the parameters from Table~\ref{tab: observation-ngc}. The green cross points the centre of the galaxy and the green circle has a radius of 5.91 kpc, i.e. the galactocentric distance of the GC. One can notice extended tidal tails of sizes greater than 1 kpc in roughly 3.66 Gyr, which is well below the age of NGC~6397 (12.87 Gyr). Consequently, we predict that NGC~6397 should exhibit tidal tails and these should be clearly observable. }
\label{fig: gc-shots}
\end{figure*}

\begin{figure*}
\centering
\includegraphics[width=0.9\hsize]{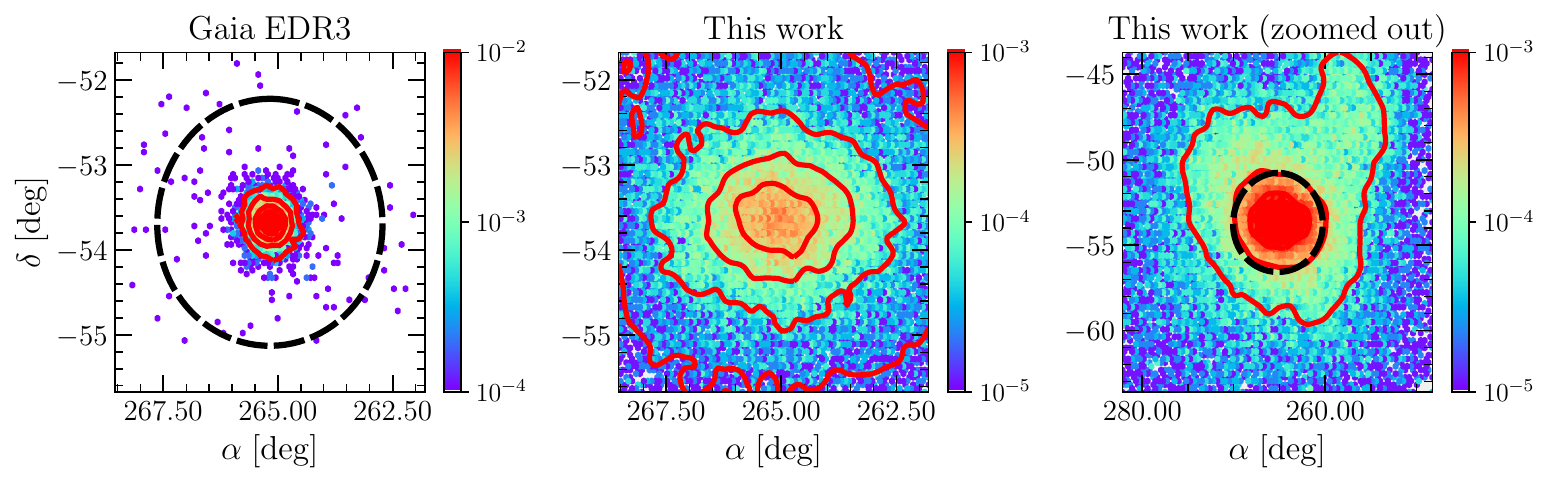}
\caption{{\it Absence of a diffuse and distorted stellar distribution in observations:} Sky plots, colour coded by the normalized star counts per bin, with respect to the total number of stars. The first column represents the filtered \textsc{Gaia EDR3} data, while the last two columns show the simulated GC in Run 3 at $T_{\rm d}^{\rm c}=$ 10 Gyr, placed at the same distance than NGC~6397. The middle column is centered in a similar box than the \textsc{Gaia EDR3} data while the last column is zoomed out in order to better distinguish the inner tidal tails of our simulations. We draw red iso-contours corresponding to 1, 1.5, 2, 2.5 and 3-$\sigma$ regions around the maximum normalized star counts and we plot a dashed black circle which represents the extent of the cluster tidal radius in each case. The amount of stars displayed in the first, second and third plots is 10460, 42840 and 81886 respectively. It shows a very concentrated distribution of stars in NGC~6397, in contrast with a strongly diffused profile of our simulated GC. The latter exhibits an extended and distorted stellar distribution in its inner part due to the strong MW tidal field, whereas NGC~6397 seems to maintain a spherical shape.}
\label{fig: tails-proj}
\end{figure*}

\begin{figure*}
\centering
\includegraphics[width=0.9\hsize]{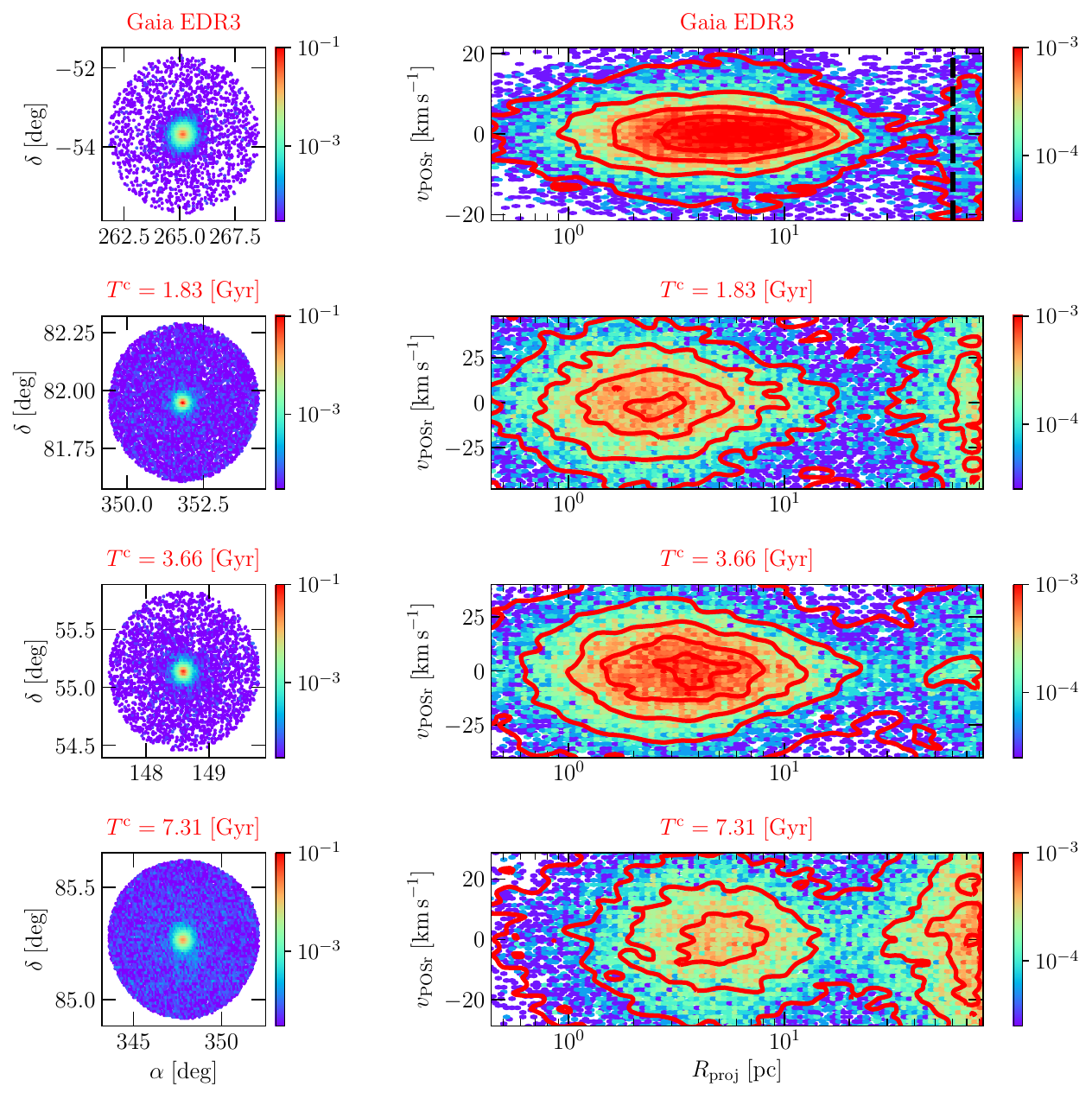}
\caption{{\it Discontinuity in density for NGC~6397 from \textsc{Gaia EDR3} data:} Comparison of \textsc{Gaia EDR3} (first row) and the GC from our Run 3, at different times (other rows). \textit{Left:} Projected sky plots. \textit{Right:} Radial direction of the proper motion as a function of the logarithmic binned projected distance along with the tidal radius in dashed black. The colour bars indicate the normalized star counts per bin, with respect to the total number of stars. The amount of stars in all plots (along with MW contaminants) is 35974, and the amount of stars beyond the tidal radius shown in the images is 790 for the observed cluster. During the formation of tidal tails for our simulated GC, we observed that a continuity in density, for the radial velocity along the projected radius, emerges below the tidal radius of approximately 100 pc. It shows the presence of potential escapers due to MW tidal effects. In contrast, for NGC~6397, there is a clear cut-off at roughly 30 pc between GC stars and MW interlopers. Indeed, the density transition in observed data seems to correspond to that of a cluster at the very beginning of its evolution in the MW tidal field as the simulated cluster at $T^{\rm c}=$ 1.83 Gyr. We conclude that there is no obvious sign of an ongoing intense tidal disruption for NGC~6397 stars.}
\label{fig: tails-velocity}
\end{figure*}

\begin{figure}
\centering
\includegraphics[width=\hsize]{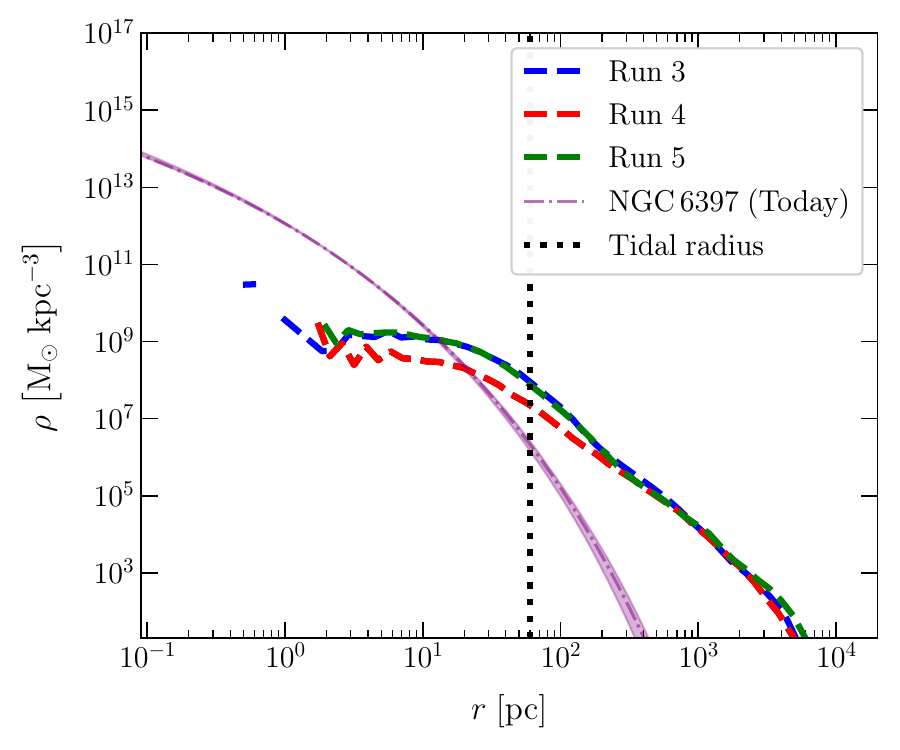}
\caption{{\it Inner stellar distribution:} Density profile of NGC~6397 at the end of the simulation, shown in dashed lines for all the models which survived for more than $T_{\rm d}^{\rm c} = 10$ Gyr. The tidal radius of NGC~6397 is shown in the vertical black dotted line while the density profile of the cluster, derived in \protect\cite{Vitral&Mamon21}, is displayed in purple dot-dashed. The number of stars beyond the tidal radius for runs 3, 4 and 5 are 60485, 51449 and 66444 for the final density profiles, respectively. For all the different simulations, the final profile is much more diffuse than the observed profile, extending themselves beyond the tidal radius of NGC~6397. According to a S\'ersic profile fit of our stripped simulated clusters, we found a S\'ersic radius $R_{\rm e}$ about $\sim$50 times greater than the one observed with Gaia data ($R_{\rm e}=$ 6.5 pc). This implies that the tidal disruption seen in our simulations is clearly more advanced than any possible tidal disruption that NGC~6397 has suffered.}
\label{fig: densities}
\end{figure}

\section{Results}

Next, we present and discuss our simulation results. We calculate the distance between the GC mass centre and MW centre at each snapshot, in order to get the orbital radius. To estimate the GC mass loss, we count only bound particles. We follow the iterative method of \cite{Baumgardt2003} to determine the GC mass over time and we use our simulation labelled as Run 3 as a standard model to present our main results.

\subsection{Tidal disruption}

We have considered the evolution of a live GC with a mass resolution of 10 M$_{\sun}$ in a MW static potential over 10 Gyr. Details of all our scenarios are given in Table~\ref{tab: n-body-models}. The disruption times were defined to be the time when 95$\%$ of the mass was lost from the GC. Table~\ref{tab: n-body-models} shows that it is possible to adjust initial parameters, such as the mass $M_{\mathrm{GC}}$, the S\'ersic radius $R_{\mathrm{e}}$ and index $n$, which ensures the survival of the cluster NGC 6397 in spite of the strong MW tidal field, as previously demonstrated by \cite{Baumgardt2003} and \cite{Shin+13}. In addition to the fact that the initial GC mass is higher, we notice that its lifetime mostly depends on its initial density profile (see Table~\ref{tab: n-body-models}). 

We also investigate the dynamics of a NGC 6397-like GC with a $10^3$ M$_{\odot}$ BH at its centre (Run 2). This run is motivated by the intermediate-mass black hole fits of 600 M$_{\odot}$ and 511 M$_{\odot}$ in the centre of NGC~6397 (\citealt{Kamann+16} and \citealt{Vitral&Mamon21}, respectively). As shown by \cite{Baumgardt2004}, a central BH in a GC will increase the escape rate of stars and thus reduce the GC lifetime. We emphasize that this effect is well observed in our simulation as reflected by the smaller disruption time of Run 2 in Table~\ref{tab: n-body-models}. 

An expected observable feature of GCs, which are orbiting within the MW, especially in the inner region, is the development of stellar tidal tails \citep{Grillmair1995,Dehnen2004}. In all our simulations, we observe the formation of these stellar structures, which originate from the epicyclic motions of a continuous stream of stars escaping the clusters \citep{Kupper2010,Kupper2012}, as the orbits of our GCs are not very eccentric ($e\sim$ 0.25). We estimate the time of formation of tidal tails for all our runs according to Section~\ref{sec: Nbody}. Our results point out that the stellar streams of our GC are formed in the first third of the cluster's life (see Table~\ref{tab: n-body-models}). Figure~\ref{fig: gc-shots} illustrates the formation of tidal tails around the GC in Run 3 (see Table~\ref{tab: n-body-models}). According to this Figure, our simulation indicates extended tidal tails of size greater than 1 kpc in roughly $T_{\rm tails}^{\rm sim}=0.5$ Gyr, which would correspond to a real time of about $T_{\rm tails}^{\rm c}=3.66$ Gyr when applying the correction factor $C_{\mathrm{t}}$ from equation~\eqref{eq: correction}. Even though the corrected time of 3.66 Gyr is just an approximation, it is well below the age of NGC~6397 (i.e. 12.87 Gyr according to \citealt{MarinFranch+09}). Consequently, we predict that NGC 6397 should exhibit tidal tails and these should be clearly observable. This result is in agreement with \cite{Montuori+07}, who simulated globular clusters, modelled by a King profile, in very strong MW tidal fields ($R_{\mathrm{GC}}<$ 4 kpc), and found tidal arms elongated for more than 1.5 kpc in less than 1 Gyr of evolution.

\subsection{Comparison between simulations and observations}

Now, we try to find any possible indication of tidal features, such as a stellar diffuse inner profile or tidal tails for NGC~6397 by comparing \textsc{Gaia EDR3} with our simulated GCs. This comparison is done in two steps. First, in Figure~\ref{fig: tails-proj}, we investigate the inner profile in the sky plot of observed and simulated clusters. Second, in Figure~\ref{fig: tails-velocity}, we analyze the outer regions of the clusters by considering the projected radial velocity as a function of the projected distance to the cluster centre. We describe these two steps in the following subsections.

\subsubsection{Sky plot}

In order to compare our simulation results with NGC~6397 using \textsc{Gaia EDR3} data, we need to project our simulated GCs on the plane of sky, which was done using the \textsc{Astropy} Python package (\citealt{AstropyCollaboration+13}), assuming the observer is located at the Sun's position. Figure~\ref{fig: tails-proj} was then constructed by projecting our simulations in the plane of sky, and in this particular case, placing the GC at the same galactic position as observed today. This allows us to check if an effect of projection or an artifact of the close position of the cluster could blur the diffuse spatial distribution of our simulated GCs. 

The first column in Figure~\ref{fig: tails-proj} shows the observed GC, without interlopers, which was done after selecting the GC stars by filtering \textsc{Gaia EDR3} with \textsc{BALRoGO}\footnote{\url{https://gitlab.com/eduardo-vitral/balrogo}} \citep{Vitral21}. We did not consider membership probabilities related to the projected distance (apart from allowing stars up to two degrees from the cluster centre) since one of our goals is to detect possible GC escapers, and therefore, our filters were based on proper motion and CMD regions only. We were also forced to keep stars with a low proper motion error (defined as in \citealt{Lindegren+18}) of half the velocity dispersion of the cluster, so the number of MW contaminants was reduced, which in turn led us to a cutting $G_{\rm mag}$ of roughly 20 mag.

Figure~\ref{fig: tails-proj} clearly shows a very concentrated distribution of stars in NGC~6397 \textsc{Gaia EDR3} data, in contrast with a strongly diffused profile of our GC in Run 3 at $T_{\rm d}^{\rm c}=$ 10 Gyr. Although the tidal tails of the simulation seem too large to be perceived in a two degree sky plot, the inner shape of observed and simulated clusters are radically different. Our simulated GC exhibits an extended and distorted stellar distribution in its inner part due to the strong MW tidal field, whereas NGC~6397 seems to maintain a spherical compact shape.

\subsubsection{Projected radial velocity}

In Figure~\ref{fig: tails-velocity}, we again compare \textsc{Gaia EDR3} data (first row) with subsequent snapshots of our GC from Run 3, drowned in mock MW contaminants and projected in the plane of sky, but keeping the original distance to the Sun from each step of the simulation. The cut in \textsc{Gaia EDR3} data was much less conservative this time, since we allowed all \textsc{Gaia EDR3} stars within a 5-$\sigma$ region around the bulk proper motion of the GC, which ensures that we do not miss possible GC stars which are part of tidal tails. This was done by accepting all stars satisfying:

\begin{equation}
    \sqrt{(\bar{\mu}_{\alpha*, \rm GC} - \mu_{\alpha*, i})^2 + (\bar{\mu}_{\delta, \rm GC} - \mu_{\delta*, i})^2} \leq 5 \, \sigma_{\rm GC},
    \label{eq: pm-check}
\end{equation}
where $\bar{\mu}_{\alpha*, \rm GC}$ and $\bar{\mu}_{\delta, \rm GC}$ are the mean proper motions of NGC~6397 in right ascension and declination, respectively, derived according to \cite{Vitral21}, while $\mu_{\alpha*, i}$ and $\mu_{\delta, i}$ are the corresponding proper motions from the \textsc{Gaia} stars. In order to account for the uncertainties on the proper motions, we also added \textsc{Gaia}-like errors to our simulated cluster and interlopers according to the recipe from \cite{Vitral21}. The selection of cluster members by considering fiducial regions on the proper motion space around the cluster bulk proper motion is a method largely employed in the literature, and have been extensively used in Gaia data (e.g., \citealt{GaiaCollaboration+18H}; \citealt{Vasiliev2019}; \citealt{Baumgardt2019}; \citealt{Sollima20}; \citealt{Kundu+20} and more recently, \citealt{Vasiliev&Baumgardt&Baumgardt21} and \citealt{Vitral21}).

The fact that we employ no other data cleaning, nor exclude stars due to their positions on the sky or in the CMD surely allows for many interlopers. These contaminants can affect the distinction between GC stars and field stars, which can naturally blur imprints of tidal tails. In order to evaluate the impact of these contaminants, we created mock MW interlopers uniformly distributed in the sky according to Appendix~\ref{app: mock-MW}, and added them to our simulated data. We decided the 
position and the extent of the field stars proper motion distribution, which follows a Pearson VII symmetric distribution, by mimicking the field stars around NGC~6397.

Figure~\ref{fig: tails-velocity} clearly demonstrates that, during the formation of tidal tails for our simulated GC, a continuity in density for the projected radial proper motion along the projected radius emerges below the tidal radius of approximately 100 pc. It shows the presence of potential escapers due to MW tidal effects. In contrast, for NGC~6397, there is a clear cut-off at roughly 30 pc between GC stars and MW interlopers, and thus, no obvious tidal tails. Indeed, the density transition in the observed data seems to correspond to that of a cluster at the very beginning of its evolution in the MW tidal field, as the simulated cluster at $T^{\rm c}=$ 1.83 Gyr. Interestingly, the stellar streams detected recently by \cite{Ibata+20} seem to have a length comparable to our simulation at this stage of the cluster's evolution (see Figure~\ref{fig: gc-shots}), even when taking into account projection effects as further explained in section~\ref{ssec: proj-effects}.

According to \textsc{Gaia EDR3} data, we state that there is no obvious sign of an ongoing intense tidal disruption for this cluster. We stress that we do manage, in spite of the MW contamination, to detect these tidal features of our simulated GC by looking at the projected radial velocity as a function of the projected distance.

\subsection{Surface density profile}

Finally, we compare the density profiles at the end of our simulations which survived for more than $T_{\rm d}^{\rm c} = 10$ Gyr, with the density profile of NGC~6397, derived in \cite{Vitral&Mamon21}, in Figure~\ref{fig: densities}. We constructed these profiles by binning the simulated GCs with 50 log-spaced radial bins and considering a stellar mass of $10$ M$_{\odot}$. One can see that for all the different simulations, the final profile (dashed) is much more diffuse than the observed profile, extending beyond the tidal radius of NGC~6397. 
Moreover, both Kolmogorov-Smirnov and Wilcoxon tests indicate strong confidence that the distributions are not the same, with respective $p-$values reaching an underflow (i.e. converging towards zero). We also fitted a S\'ersic profile to our stripped simulated clusters depicted in Figure ~\ref{fig: densities} and we found a S\'ersic radius $R_{\rm e}$ above 50 times greater than the value presented in Table~\ref{tab: observation-ngc}. This indicates that the tidal disruption seen in our simulations is more advanced than any possible tidal disruption that NGC~6397 has suffered.
 
\section{Discussion}
\label{sec: discussion}

Our observational data analysis establishes that several indicators such as the stellar distribution and the radial velocity along the projected radius of NGC~6397 seem to rule out the presence of extended tidal tails, contrary to the predictions of our simulations. Here, we provide and evaluate several solutions, which could explain the discrepancy between our simulations and the \textsc{Gaia EDR3} observations of NGC~6397.

\subsection{Projection effects}
\label{ssec: proj-effects}

\begin{figure}
\centering
\includegraphics[width=0.47\textwidth]{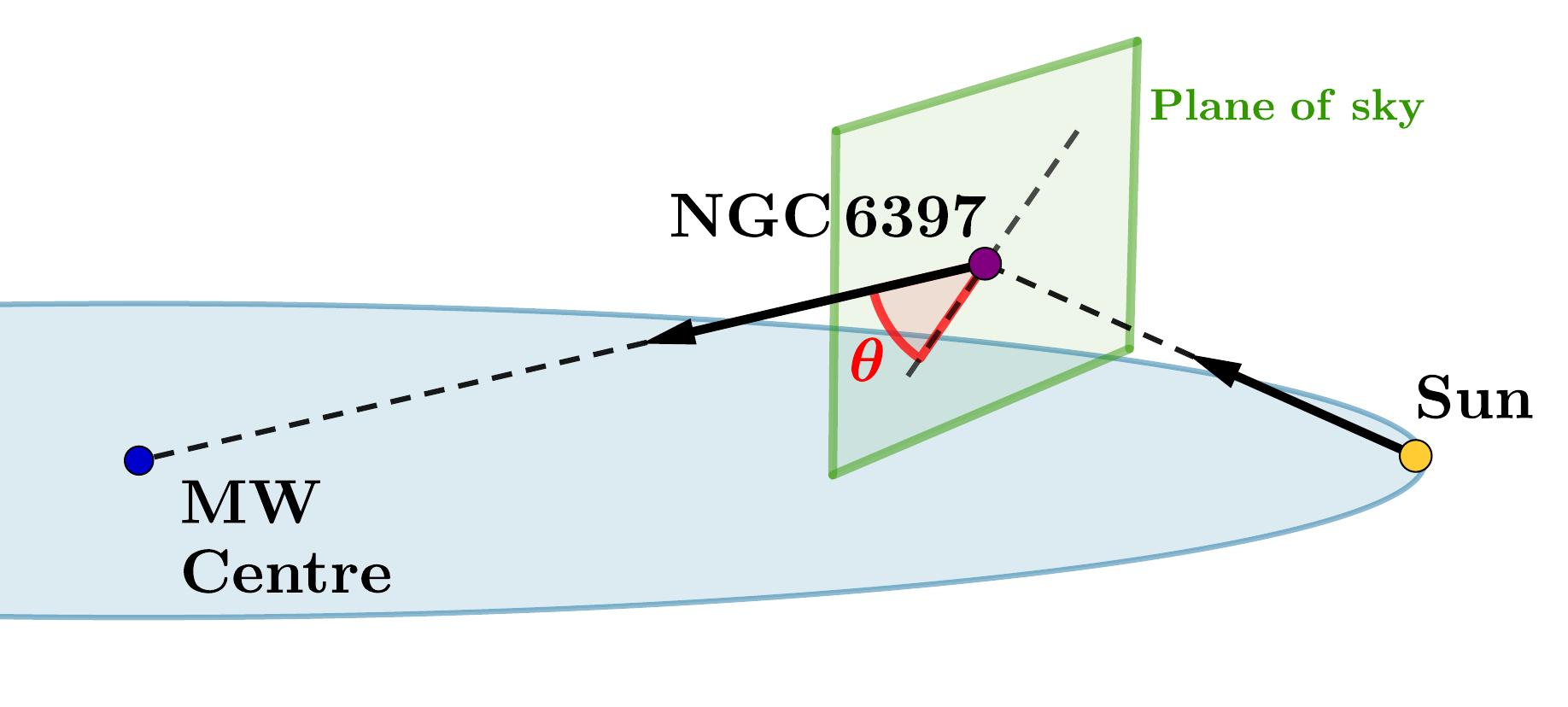}
\caption{{\it Geometry of tails:} We represent the projection effects related to tidal tails pointing towards the galactic centre, with $\theta$ being the angle between the galactocentric direction and the plane of sky centered on NGC~6397. The positions of the sources are not in agreement with their true positions, for better visualization.}
\label{fig: tails-gemoetry}
\end{figure}

\begin{figure*}
\centering
\includegraphics[width=0.9\hsize]{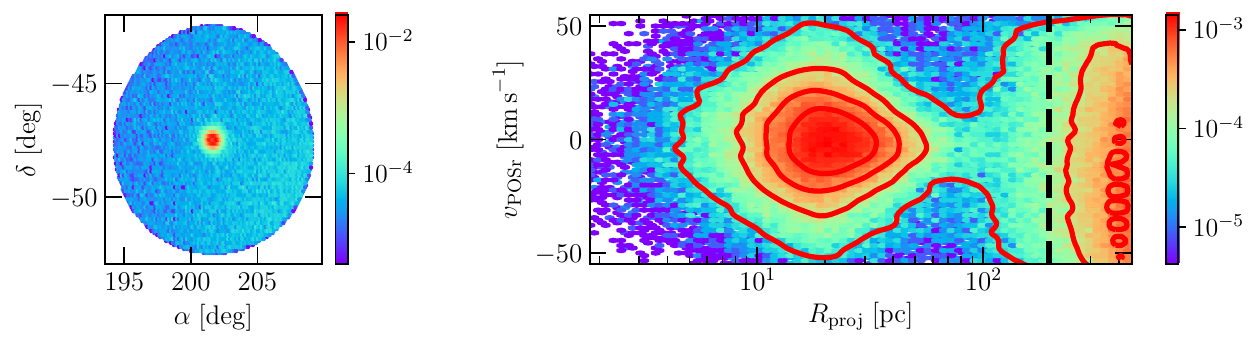}
\includegraphics[width=0.9\hsize]{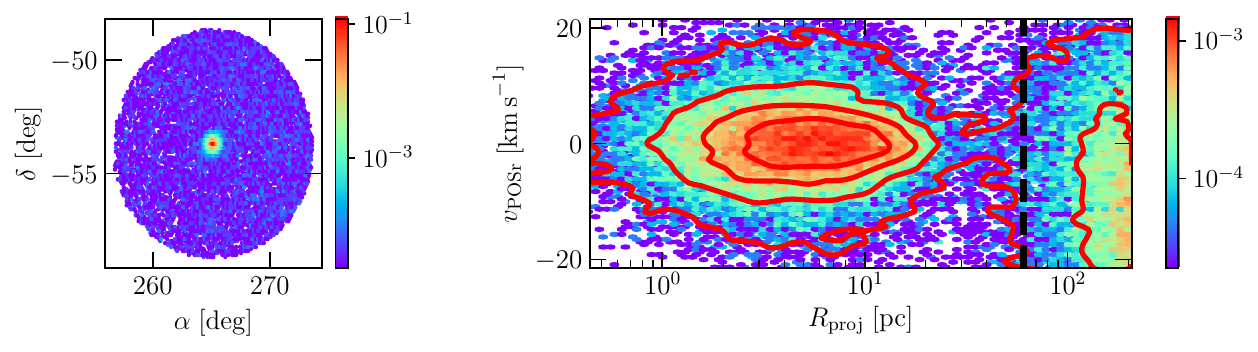}
\caption{\textit{Comparison with $\omega$ Cen:} Analysis of the globular cluster $\omega$ Cen, similar to Figure~\ref{fig: tails-velocity}, and comparison with NGC~6397. The \textsc{Gaia EDR3} data, this time observed in a 5-degree cone search, which explains the high number of stars at high projected radii. \textit{Left:} Projected sky plot of the GC. \textit{Right:} Radial direction of the proper motion of the globular cluster as a function of the logarithmic binned projected distance along with the tidal radius of each case in dashed black. The colour bars indicate the normalized star counts per bin, with respect to the total number of stars. We used a distance to the Sun of 5.24 kpc and a tidal radius of 196.46 pc for $\omega$ Cen \protect\citep{Baumgardt2019}. The upper plot presents 237692 stars, among which 67639 lie beyond the tidal radius of NGC~5139 and the lower plot displays 44599 stars, among which 9415 lie beyond the tidal radius of NGC~6397. As in our simulations in Figure~\ref{fig: tails-velocity}, $\omega$ Cen exhibits the continuity in density, for the radial velocity along the projected radius, which indicates the presence of tidal arms. For the observed NGC~6397 extending up to 5 degrees, there is still a clear cut-off and no sign of tidal arms, but only of a stronger MW contamination.}
\label{fig: omeg-cen}
\end{figure*}

One of the possible explanations for the lack of tails in the observations could be a projection effect, since the tails could be aligned along the line-of-sight connecting the Sun and NGC~6397. If this is the case, the projection of these tails would be negligible on the sky plot of Figure~\ref{fig: tails-proj}, as well as in the outer projected radii of Figure~\ref{fig: tails-velocity}. 

In order to test this assumption, we refer to the work of \cite{Klimentowski+09} and \cite{Montuori+07}, who analyzed the orientation of tidal tails in dwarf spheroidal galaxies and GCs, respectively, by studying the outcome of $N-$body simulations. Both studies report that in the vicinity of the dwarf or the GC, the tails are typically oriented towards the MW and not along the orbit (see Figures~2 and 3 from \citealt{Montuori+07}). Indeed, it is in the vicinity of those sources (i.e. at distances closer than $\sim$1 kpc) that the tails are denser and therefore most likely to be detectable. This is well suited for our analysis as we restrict our study to only two (five, in Figure~\ref{fig: omeg-cen}) degrees around the cluster centre.

Therefore, we verify if the sky projection of the direction connecting the NGC~6397 and the MW centre is washed out due to an alignment problem (i.e. $\theta \approx 90$ degrees in Figure~\ref{fig: tails-gemoetry}). We derive $\theta = 57.43$ degrees, which is sufficient for us to detect extended tidal tails, since the projection of such components would yield a length of roughly $L \times \cos{\theta} = L \times 0.54$, with $L$ being the length of the tidal tails in the direction of the MW centre (see Figure~\ref{fig: tails-gemoetry}). the large projection of the tails in the sky derived above is then an important argument for this bias to be neglected.

\subsection{Limiting magnitude}

\textsc{Gaia EDR3}, as well as its predecessor, has a magnitude limit of $G_{\rm mag} \sim 20$, which basically means that the faintest GC stars, with masses $\lesssim 0.5$ M$_{\odot}$, are not taken into account. Naturally, fainter stars tend to occupy distances farther from the GC centre than brighter stars, due to mass segregation (\citealt{Binney2008}), and such a relation has been robustly observed by \cite{Heyl+12} and \cite{Vitral&Mamon21} for NGC~6397. Therefore, one possibility is that NGC~6397 might have tidal tails which are just not detected by \textsc{Gaia EDR3}, as a consequence of how deep, in magnitude, its observations can probe. However, \cite{Vitral&Mamon21} found that, although the fainter stars do follow a more diffuse distribution in space, their radial extent is only about two times greater than the brighter stars (i.e. $R_{\rm e}=6.5$ pc, see Table~\ref{tab: observation-ngc} for a comparison), which remains a small radial extent when compared to our simulated GC from Run 3 presenting a S\'ersic radius of $R_{\rm e}\sim$ 200 pc and tidal tails of more than 1 kpc long in early stages of the cluster life. 

\subsubsection{Comparison with other globular clusters}

We now test if this limiting magnitude could affect, for example, Figures~\ref{fig: tails-proj} and \ref{fig: tails-velocity}, by blurring or erasing the contribution of faint stars in tidal tails drowned in MW contaminants. For that, we generate plots similar to Figure~\ref{fig: tails-velocity} for GCs known to have obvious tidal tails, such as Pal 5 (\citealt{Odenkirchen+01}) and NGC~5139 ($\omega$ Cen, \citealt{Ibata+19}).

Indeed, our method was not able to detect strong imprints of the massive tidal tails of Pal 5, given that it is located at roughly 21.6 kpc away from the Sun. As Pal 5 is about nine times more distant than NGC~6397, the limiting magnitude of \textsc{Gaia EDR3} prevents us from observing many of the faint stars that dominate the tails of the former cluster, whose apparent magnitude depends on their distance to us. Therefore, we decide to test our method with a closer GC, $\omega$ Cen, known to have important tidal tails (e.g. \citealt{Ibata+19} and \citealt{Sollima20}), and whose distance and MW contamination are not too different from NGC~6397 (the values of distance to the Sun and galactocentric distance of $\omega$ Cen are 5.2 kpc and 6.4 kpc, respectively, \citealt{Harris10}). Since NGC~5139 is still more distant than NGC~6397, detecting any tidal imprints around it through our method from Figure~\ref{fig: tails-velocity} would indicate the reliability of this procedure.

The first row of Figure~\ref{fig: omeg-cen}, which also selects all \textsc{Gaia EDR3} stars within a 5-$\sigma$ proper motion region, this time inside a five-degree cone search around $\omega$ Cen, shows that although the MW contamination does play a role in blurring the GC escapers, we are still able to observe the continuity in density, for the proper motion radial component along the projected radius, which indicates the presence of tidal arms. Even if \citealt{Montuori+07} pointed out that $\omega$ Cen has tidal tails deviating from the galactocentric direction and extending toward the galactic plane, our method highlights the external structure. Besides, $\omega$ Cen in Figure~\ref{fig: omeg-cen} is similar to second and third snapshots of Figure~\ref{fig: tails-velocity}. We display, in the second row, the similar plot of NGC~6397 for comparison. Thus, if ever NGC~6397 has extended tidal tails formed by fainter stars blurred by MW contamination, it would be curious if such an effect is not repeated for $\omega$ Cen.

For the observed NGC~6397 extending up to 5 degrees, which is the aperture usually probed in this kind of study \citep{Sollima20}, there is still a clear cut-off and no sign of tidal arms, but only of a stronger MW contamination at outer radii (see Figure~\ref{fig: omeg-cen}). Indeed, \citealt{Klimentowski+09} stress that the most evident signs of tidal structures will be seen closer to the analyzed source, where the stars are more clumped together.

\subsection{Dark matter profile of the MW}

In our MW mass model, we have assumed a NFW profile for the DM halo. However, an evidence for a DM core in the MW was claimed by \cite{Portail2017} in order to explain the high velocity dispersion of the stars in the Galactic bulge. They predicted a small DM core of few kpc in size. In all runs, the GC orbits around the MW centre at a distance between 5.1 and 9.2 kpc. As the GC does not orbit in the regions where a DM core can be produced by feedback \citep{Lazar2020}, it will not feel the dynamical impact of a different DM density profile for the MW. Moreover, we emphasize that the gravitational potential is dominated by stars of the bulge in this inner region of the MW. Therefore, varying the enclosed mass in the inner region of the MW is not a viable solution to efficiently delay the formation of tidal tails around the GC in order to reproduce the \textsc{Gaia EDR3} observations.

\subsection{GCs embedded in dark matter}

We now discuss the possible influence of the presence of DM in GCs during their evolution. It has also been proposed that GCs may have a galactic origin, where GCs are formed within DM minihalos in the early Universe (e.g. \citealt{Peebles1984,Bromm2002,Mashchenko&Sills&Sills05,Ricotti2016}). Then, these GCs could have merged to later become a part of the present-day host galaxy.

Until now, these DM halos have not been detected. More precisely, it was pointed out that the ratio of the mass in DM to stars in several GCs is less than unity \citep{Shin+13,Conroy2011,Ibata+13,Moore1996,Baumgardt2009,Lane2010,Feng2012,Hurst2015}. Even if GCs are proven not to have a significant amount of DM, it does not preclude them from having been formed originally within a DM minihalo. A natural explanation is that they have lost their DM over time. Indeed, there are several internal dynamical processes which could eject DM from GCs such as DM decay \citep{Peter2010} and feedback processes \citep{Pontzen2012,Davis2014}. It was also shown that GCs orbiting in the inner regions of their host galaxies may lose a large fraction of their primordial DM minihalo due to tidal stripping (e.g. \citealt{Bromm2002,Mashchenko&Sills&Sills05,Saitoh2006,Bekki2012}). That is the reason why the main mechanism by which our GC could have lost its DM minihalo is through severe tidal interactions with our Galaxy given its current position. For instance, it was demonstrated that if the Fornax GCs were embedded in DM, they would have lost a large fraction of their DM minihalos, whose masses would have been in good agreement with the constraints for GCs \citep{Boldrini2020a}. Nevertheless, GCs at a large distance from the MW centre could have retained a significant fraction of DM because it was not completely stripped by the Galaxy. Even if observations of these GCs such as NGC 2419 and MGC1 highlight that they do not possess significant DM today \citep{Conroy2011,Ibata+13}, it does not exclude the existence of DM in GCs but suggest that there is not necessarily a unique formation mechanism for GCs in general. 

Contrary to our simulated GCs, which exhibit obvious stellar tidal tails, which are well extended in the sky, we argue that NGC 6397 does not possess such strong tidal imprints, based on \textsc{Gaia EDR3} data (see Figures~\ref{fig: tails-velocity} and~\ref{fig: omeg-cen}). It was claimed that these tidal structures should not form if they were embedded in DM minihalos \citep{Grillmair1995,Moore1996,Mashchenko&Sills&Sills05,Odenkirchen2003}. Indeed, adding this dark component to this GC could help to avoid the formation of stellar tidal tails. Moreover, it was shown that DM minihalos following a core profile are much easier to disrupt tidally in galaxy than minihalos described by an NFW profile \citep{Mashchenko&Sills&Sills05}. A GC embedded in NFW DM minihalo could keep its stellar structure almost intact but be still dominated by DM in its outskirts \citep{Heggie&Hut96,Baumgardt2009,Lane2010,Penarrubia2017}. It could be possible to detect the remnant of the DM halo by looking at the projected velocity dispersion profiles of stars \citep{Penarrubia2017}. Indeed, the presence of DM will manifest itself in the form of a heating of the stars beyond the tidal radius. Figure~6 from \citealt{Vitral&Mamon21} shows a clear increase in the dispersion profile of NGC~6397 beginning at roughly 17 arcmin (i.e. 20 per cent of the tidal radius).

We pointed out that \textsc{Gaia EDR3} data reveals that stars in the central region of NGC~6397 seem to follow a spherical distribution (see Figure~\ref{fig: tails-proj}). Curiously, it was argued that a stellar cluster evolving inside a DM minihalo would have close to this sort of distribution in its denser part, where the stars dominate over DM \citep{Mashchenko&Sills&Sills05}. We stress that DM provides a potential solution to the case of NGC~6397. A DM minihalo could be responsible for both the inner spherical shape and the absence of tidal tails in this cluster, as the stellar part of a GC embedded in DM would be more resilient to tidal disruption. 

\subsection{Ex-situ origin}

It was also suggested that nuclear star clusters of tidally-stripped galaxies may be the progenitors of GCs, especially massive ones such as G1 and $\omega$ Cen \citep{Freeman1993,Bekki2003,Bekki2012,Boker2008,Meylan2001}. They are the most massive clusters of their parent galaxies, M31 and the MW, respectively. Nucleated dwarf galaxies can be transformed into GCs due to tidal stripping of the dwarfs by the strong gravitational field of galaxies \citep{Bekki2002,Bekki2003}. 
It is important to note that this scenario is also motivated by the presence of a far more complex stellar population in clusters such as $\omega$ Cen (three distinct populations, \citealt{Pancino2003}), compared to normal clusters. NGC 6397 hosts only two stellar populations \citep{Milone+12b}.

Furthermore, these GC-like systems were embedded in DM halos \citep{Taylor05,Boker2004,Walcher2005,Walcher2006}. GCs originated from nucleated dwarfs have therefore formed outside the central regions of the MW. That is the reason why these GC-like systems can have no DM at the present day. In fact, most of the initial DM halo was stripped away from the GC during the first several orbits, even if they are massive at their birth \citep{Wirth2020}. 

The fact that a GC may be a nuclear star cluster remnant could explain the absence of its stripping in the dense MW central region, but it is also used to argue that it may have an intermediate-mass black hole (IMBH) at its centre \citep{Bahcall1976}. Indeed, it was claimed that G1 and $\omega$ Cen could contain IMBHs with a mass of $10^4$ M$_{\odot}$ \citep{Gebhardt2002,Gebhardt2005,VDM2010}. However, it was recently demonstrated that the presence of a central IMBH in NGC 6397 was ruled out, in favour of a diffuse dark inner sub-cluster of stellar remnants \citep{Vitral&Mamon21}, well explained by a strong concentration of faint white dwarfs \citep{Kremer+21}.

\section{Conclusion}

In this work, we performed $N$-body simulations on GPU in an attempt to reproduce the past 10 Gyr of dynamical evolution of a globular cluster presenting similar characteristics to the nearby globular cluster NGC~6397, in the strong tidal field of a Milky Way-like galaxy. We considered a purely baryonic globular cluster, and we used the prescription from \citeauthor{Baumgardt2003} (\citeyear{Baumgardt2003}, see our equation~\eqref{eq: correction}, in order to access the corrected time of evolution of the cluster, while keeping a feasible mass resolution.

The intense background of dark matter and stars in our simulations is sufficient to disrupt the cluster in a Hubble time. We demonstrate that the lifetime of the simulated cluster depends mainly on its initial density profile and mass. More importantly, we find that more than 6 Gyr ago, right after the first third of the cluster's life, all of our simulated GCs presented tidal tails, which extended farther than 1 kpc long. We subsequently compare our simulations and observations of NGC~6397 from \textsc{Gaia EDR3} data. Despite its recent passage through the galactic disk (\citealt{Rees&Cudworth03}), we find that several indicators seem to rule out the presence of obvious tidal tails around NGC~6397. Since we believe that the analysis performed with \textsc{Gaia EDR3} data was robust and contradictory to our simulations, where GCs exhibit extended tails, it is much more likely that the cause of such discrepancy is due to the origin scenario of the evolving cluster, as well as some of the environmental conditions of our simulations. Indeed, we have considered three main factors that could influence the formation of tidal tails in NGC~6397: the dark matter profile of the Milky Way, an extra-galactic origin for this globular cluster, as well as their formation in dark matter minihalos.

We argue that the most likely flaw of our simulation was to consider a purely baryonic globular cluster at the beginning of the simulation. In fact, when assuming a scenario where globular clusters are formed in a dark matter minihalo (\citealt{Peebles&Dicke68,Peebles1984}), we would expect it to protect the cluster stars against intense tidal stripping, which would act as a shield and be itself stripped instead of the luminous matter. The fact that we do not manage to robustly observe considerable amounts of dark matter in globular clusters, and particularly in NGC~6397, could be attributed to such initial dark matter minihalos being stripped by the combination of internal processes of the globular cluster such as stellar evolution and supernovae, and the intense tidal field of the Milky Way.

The future releases of the \textsc{Gaia} mission will undoubtedly shed light onto this mystery since more reliable and complete astrometry can help to robustly assign cluster membership probabilities to the stars in the field of the globular cluster, and thus, better trace escaping cluster stars. It will also allow us, by targeting the cluster outskirts, to probe the velocity dispersion in and beyond the tidal radius and thus test predictions concerning the formation of globular clusters in dark matter minihalos (e.g., \citealt{Penarrubia2017}). In a future work, we intend to test this dark matter scenario in our simulations and therefore test if such initial conditions can adequately explain the absence of extended tidal tails in NGC~6397.

\section*{Acknowledgements}
We thank the reviewer for their constructive feedback which helped to improve the quality of the manuscript. We thank Holger Baumgardt, Jorge Pe\~{n}arrubia, Joseph Silk, David Valls-Gabaud and Gary Mamon for useful comments and suggestions to improve the quality of this work. We also thank Miki Yohei for providing us the non-public $N$-body code, \textsc{gothic}. We would like also to thank Dante von Einzbern for his constructive suggestions on improving the manuscript. 

\section*{Data availability}

The data that support the plots within this paper and other findings of this study are available from the corresponding author upon reasonable request.

\bibliographystyle{mnras}
\bibliography{src} 




\appendix

\section{Mock Milky Way interlopers}
\label{app: mock-MW}

In this appendix, we describe how we generated Milky Way interlopers for our simulated globular clusters in Figure~\ref{fig: tails-velocity}. The general method is described in section 2.6 of \cite{Vitral21}. In addition, we describe here how we decided the input parameters of this method such as the interlopers maximum projected radius, bulk proper motions and their number.

\subsection{Spatial distribution and proper motions}

We created stars uniformly distributed in a spherical cap of projected radius $R_{\rm simu} = R_{\rm data} \times D_{\rm simu} / D_{\rm data}$, where $R_{\rm data}$ is the maximum projected radius of our \textsc{Gaia EDR3} data in Figure~\ref{fig: tails-velocity} (i.e., 2 degrees), and $D_{\rm simu}$ and $D_{\rm data}$ are the distances of the simulated and observed clusters, respectively, with respect to an observer located at the Sun.

The number of interlopers was naively selected by taking an approximate interloper fraction in the region limited by $R_{\rm simu}$, calculated by considering the number of GC stars as the stellar count in the \textsc{Gaia} cone search minus $\pi R_{\rm simu}^2 \times \Sigma_{\rm MW}$, with $\Sigma_{\rm MW}$ (the interlopers surface density) assumed constant and estimated as the surface density of the whole data beyond twelve arcmin (about three times the S\'ersic radius). This interloper fraction was used with respect to the number of stars in the simulation having projected radius smaller than $R_{\rm simu}$. At the end, we selected randomly, among all stars in the subset, the same number of stars present in the plot of NGC~6397, so we could mimic the incompleteness of the data.

The proper motions of the field stars created above follows a Pearson VII symmetric distribution with parameters similar to the ones of NGC~6397 Milky Way interlopers. Notably, the distance of the interloper bulk proper motion was taken as $d_{\rm simu} = d_{\rm data} \times \sigma_{\rm simu} / \sigma_{\rm data}$, with $\sigma$ the one dimensional velocity dispersion in proper motions. The velocity dispersion and Pearson VII slope (see \citealt{Vitral21} for details) were chosen to be similar to the values from the interlopers around NGC~6397 (i.e., $a=5.43$ and $\tau=-6.36$).

\bsp	
\label{lastpage}
\end{document}